\documentclass[aps,pra,preprint, superscriptaddress, amsmath,amssymp,showpacs]{revtex4}

\usepackage[T1]{fontenc}
\usepackage[latin1]{inputenc}
\usepackage[dvips]{graphicx,color}
\usepackage{rotating}
\usepackage{bm}        
\usepackage{amssymb}   
\usepackage{amsmath} 
\usepackage{bm}
\usepackage[capitalize]{cleveref}

\def\prb#1#2#3{Phys.~Rev.~B~{\bf #1},\ #2\ (#3)}

\def\jcp#1#2#3{J.~Chem.~Phys.~{\bf #1},\ #2\ (#3)}

\def\pra#1#2#3{Phys.~Rev.~A~{\bf #1},\ #2\ (#3)}
\def\prl#1#2#3{Phys.~Rev.~Lett.~{\bf #1},\ #2\ (#3)}

\def\rmp#1#2#3{Rev.~Mod.~Phys.~{\bf #1},\ #2\ (#3)}

\def\k1{k_1}
\def\k2{k_2}
\def\q1{q_1}
\def\q2{q_2}

\def\({\left (}
\def\){\right )}
\def\[{\left [}
\def\]{\right ]}

\newcommand{\beq}{\begin{equation}}
\newcommand{\eeq}{\end{equation}}
\newcommand{\bra}[1]{\langle #1 |}
\newcommand{\ket}[1]{| #1 \rangle}

\newcommand{\half}{\frac{1}{2}}

\newcommand{\beqs}{\begin{subequations}}
\newcommand{\eeqs}{\end{subequations}}

\newcommand{\eqs}[1]{\cref{#1}}
\newcommand{\nbar}{\bar{n}}

\newcommand{\gbar}{\bar\gamma}
\newcommand{\rbar}{\bar{r}}
\newcommand{\Real}{\operatorname{Re}}
\newcommand{\Imag}{\operatorname{Im}}

\begin{document}
\date{\today}
\title{Quantum dynamics of incoherently driven V-type system: Analytic solutions beyond the secular approximation} 

\author{Amro Dodin}
\affiliation{Chemical Physics Theory Group, Department of Chemistry, and Center for Quantum Information and Quantum Control, University of Toronto, Toronto, Ontario, M5S 3H6, Canada}

\author{Timur V. Tscherbul}
\email{ttscherbul@unr.edu.}
\affiliation{Chemical Physics Theory Group, Department of Chemistry, and Center for Quantum Information and Quantum Control, University of Toronto, Toronto, Ontario, M5S 3H6, Canada}
\affiliation{Department of Physics, University of Nevada,  Reno, NV 89557, USA}

\author{Paul Brumer}
\affiliation{Chemical Physics Theory Group, Department of Chemistry, and Center for Quantum Information and Quantum Control, University of Toronto, Toronto, Ontario, M5S 3H6, Canada}

\begin{abstract}
We present closed-form analytic solutions to non-secular Bloch-Redfield master equations for quantum dynamics of a V-type system driven by weak coupling to a thermal bath. We focus on noise-induced Fano coherences among the excited states induced by incoherent driving of the V-system initially in the ground state. For suddenly turned-on incoherent driving, the time evolution of the coherences is determined by the damping parameter $\zeta=\frac{1}{2}(\gamma_1+\gamma_2)/\Delta_p$, where $\gamma_i$ are the radiative decay rates of the excited levels $i=1,2$, and $\Delta_p=\sqrt{\Delta^2 + (1-p^2)\gamma_1\gamma_2}$ depends on the excited-state level splitting $\Delta>0$ and the angle between the transition 
dipole moments in the energy basis.  The coherences oscillate as a function of time in the underdamped limit  ($\zeta\gg1$), approach a long-lived quasi-steady state in the overdamped limit ($\zeta\ll 1$), and display an intermediate behavior at critical damping ($\zeta= 1$). The sudden incoherent turn-on generates a mixture of excited eigenstates $|e_1\rangle$  and $|e_2\rangle$ {\it and their in-phase coherent superposition} $|\phi_+\rangle = \frac{1}{\sqrt{2\bar{r}}}(\sqrt{r_1} |e_1\rangle + \sqrt{r_2}|e_2\rangle)$, which is remarkably long-lived in the overdamped limit (where $r_1$ and $r_2$ are the incoherent pumping rates). Formation of this coherent superposition {\it enhances} the decay rate from the excited states to the ground state.  In the strongly asymmetric V-system where the coupling strengths between the ground state and the  excited states differ significantly, we identify additional asymptotic quasistationary coherences, which arise due to slow equilibration of one of the excited states.
Finally, we demonstrate that noise-induced Fano coherences are maximized with respect to populations when $r_1=r_2$ and the transition dipole moments are fully aligned.


\end{abstract}

\maketitle
\clearpage
\newpage

\section{Introduction}

Weak-field thermal excitation of multilevel quantum systems is the primary step in many physical, chemical,   and biological phenomena, ranging from photosynthesis \cite{Blankenship,FlemingARPC} to photovoltaic energy conversion \cite{PV}, visual phototransduction \cite{RickeBaylor} and nanoscale heat transfer \cite{HeatTransfer}. Several of these phenomena involve absorption of sunlight, followed by conversion of solar energy into (electro)chemical energy. Since the primary light reactions steps of photosynthesis are extremely efficient, detailed understanding of the physics of solar energy capture and transformation  can lead to new insights into the bio-inspired design of highly efficient artificial photovoltaic devices (such as solar cells and photodetectors) \cite{CreatoreChin1,CreatoreChin2,Scully11,Scully13}.
For this reason, this work discusses sunlight excitation. However, our results are relevant to excitation by any incoherent source (e.g. incoherent phonons). 
 

The primary absorption of sunlight in biological systems occurs in photosynthetic light-harvesting complexes (LHCs) which also funnel the excitation energy to the reaction center, where it is used to drive the subsequent "dark" reactions of photosynthesis \cite{Blankenship,FlemingARPC,ScholesNatChem}.  The initial stages of photosynthetic light harvesting  can be probed by initiating and measuring the dynamics of molecular excitations in real time via advanced spectroscopic techniques such as the two-dimensional photon echo spectroscopy (2DPE) \cite{FlemingARPC,Jonas}.  Recent 2DPE experiments  revealed long-lasting, wavelike dynamics of energy transfer between the different chromophores in the Fenna-Matthews-Olson (FMO) complex of green sulphur bacteria \cite{Fleming07} and in the PC645 complex of photosynthetic algae \cite{Scholes10}. These room-temperature 2DPE experiments yielded coherence lifetimes in excess of 500 fs, much longer than expected for the electronic coherences in typical condensed-phase environments at room temperature \cite{FlemingARPC}. While these surprising discoveries stimulated a wave of theoretical and experimental research into quantum coherent effects in  LHCs \cite{AureliaReview,FlemingARPC,NaturePhysicsReview}, the origin of the observed coherences (electronic vs. vibrational) and their relevance to the naturally occurring biological processes remains a subject of vigorous debate \cite{Jonas}.

The concern regarding the role of quantum coherences in solar light-harvesting is relevant because 2DPE experiments use ultrafast transform-limited light pulses to excite coherent, time-evolving superpositions of molecular eigenstates (wavepackets). In contrast, the Sun emits blackbody radiation at $T=5800$~K, which is a statistical mixture of number states \cite{Aurelia15}, characterized by a diagonal density matrix in the number-state representation \cite{MandelWolf}. Because of the absence of a phase relation between the different components of the mixture, sunlight is expected to populate molecular eigenstates, not producing any coherences between them. Using these arguments, which are in keeping with the standard Einstein theory of light-matter interaction \cite{Loudon}, several groups showed that there is no dynamical evolution of the excited states prepared by incoherent light \cite{MoshePaul,Kassal}. More recent results \cite{Grinev} do suggest, however, possible time dependence at shorter times in complex molecular systems.

The most general theory of weak-field incoherent excitation of multilevel quantum systems is based on the Bloch-Redfield (BR) quantum master equations, which  describe the time evolution of the reduced density operator of a few-level system interacting with a continuum of quantized radiation field modes  \cite{CTbook}.  The Pauli rate equations underlying the Einstein theory \cite{Loudon} follow from these equations upon neglecting the non-secular terms, which describe the coupling between the populations and  coherences. 
  The non-secular terms are responsible for the phenomenon of Fano interference between the different incoherent excitation pathways leading to the same final states \cite{Scully92,Scully06,Keitel,Agarwal74}. 
Similar coherences (albeit induced by laser excitation) give rise to a host of remarkable phenomena in quantum optics such as electromagnetically induced transparency \cite{Harris,EIT}, slow light \cite{EIT,Hau}, and lasing without inversion \cite{Harris,SZbook}. 
The existence of Fano coherences in few-level quantum optical systems driven by {\it incoherent light} \cite{Scully92,Scully06,Keitel} naturally raises the question of their relevance in the context of natural light-harvesting and artificial photovoltaics, where they have been proposed to enhance the efficiency of quantum heat engines \cite{Scully11,Scully13}. However, despite early recognition of these noise-induced coherences  \cite{Scully92}, their properties  remain incompletely understood.

Previous theoretical work has focused on steady-state Fano coherences in model V- (or $\Lambda$)-type systems with degenerate upper (or lower) levels \cite{Keitel}. In particular, Kozlov {\it et al.} demonstrated that incoherent excitation can lead to non-vanishing steady-state coherence between the degenerate upper levels  \cite{Scully06}.  Agarwal and Menon \cite{Agarwal01} showed that the noise-induced Fano coherences between non-degenerate eigenstates vanish in the long-time limit, leading to a canonical steady state. They also identified the signatures of the coherences in steady-state fluorescence spectra. 
 Hegerfeldt and Plenio demonstrated that Fano coherences can give rise to quantum beats and to an extended dark period in time-resolved fluorescence emitted by V- and $\Lambda$-type systems excited by incoherent light \cite{Plenio93}, while Altenm{\"u}ller clarified the origin of these quantum beats as originating from a factorized system-reservoir initial state  \cite{Altenmuller}. Similar results have been obtained for the $\Lambda$-type system \cite{Agarwal99,Keitel,Ou}, where the coherence can be generated by spontaneous decay to  closely spaced  ground levels.

 We have recently provided analytical solutions to the BR equations  {\it for an arbitrary excited-state spacing $\Delta$}  and identified two important regimes of incoherent excitation for the specific case of equal pumping rates \cite{prl14}. In the large-spacing regime $\Delta/\gamma \gg1$, the coherences between the excited eigenstates display an oscillating behavior, decaying on the timescale $\tau_s=1/\gamma$, where $\gamma$ is the spontaneous decay rate of the excited states. In the opposite regime of closely spaced excited levels $\Delta/\gamma \ll 1$, the coherences decay on the timescale $\tau_\Delta = \gamma/(2\Delta^2)$, thus suggesting that near-degenerate energy levels can maintain coherence on a very long timescale \cite{prl14}.
However, this preliminary work did not consider several factors 
that affect the coherence dynamics: (1) the relative magnitude 
of incoherent pumping rates, and (2) the magnitude of the 
population-to-coherence coupling (or the transition dipole 
moment alignment factor 
$p=\bm{\mu}_{ge_1}\cdot\bm{\mu}_{ge_2}/(|\mu_{ge_1}||\mu_{ge_2}|$). Here
$\bm{\mu}_{ge_i}$ are the transition dipole moments between the 
ground and $i^{th}$ excited state \cite{Scully06,Keitel,prl14}. While it is known that the Fano coherence gets smaller with decreasing $p$ and eventually vanishes at $p=0$, the exact dependence on $p$ remains unexplored, as is the case of unequal pumping rates. Investigating noise-induced coherent dynamics  away from the  $p=1$ limit explored previously \cite{prl14,Scully06,Scully11,Scully13} is particularly important for molecular systems, where transition dipole moments will typically have a broad range of magnitudes and orientations.

Below, we present analytic solutions to the BR equations of motion describing incoherent excitation of the V-system for arbitrary values of  pumping rates and $p$. Extending our previous work \cite{prl14}, we find that in the weak-pump limit the solutions can be classified in terms of two parameters defined in analogy to the damped harmonic oscillator: (1) $\zeta$, which characterizes the dynamics of the excited state coherences in direct analogy to the damping coefficient of a classical oscillator, and (2) $\eta$, which quantifies the anharmonicity of the coherent dynamics.  We show that the relevant parameter, which governs the dynamical evolution of noise-induced coherences in the asymmetric V-system is the  ratio $\zeta^{-1}=\Delta_p/\gamma$ of ``renormalized'' excited-state splitting  $\Delta_p = \sqrt{\Delta^2+(1-p^2)\gamma_1\gamma_2}$ to the average decay width $\gamma$. We identify two key dynamical regimes: underdamped ($\zeta \ll 1$), where the excited-state coherences oscillate multiple times before decaying to zero on the timescale $\tau_s=1/\bar{\gamma}$, and overdamped ($\zeta \ll 1$), where the coherences approach a quasisteady state after a short time interval ($\tau_s)$  following the initial turn-on of  incoherent driving, and then stay constant for $\tau_{long}=\gamma/\Delta_p^2$. In this regime, coherences  can be made arbitrarily long-lived by reducing $\Delta_p$. A critical damping regime occurs at $\zeta=1$ , where the coherences display an intermediate behavior  between the overdamped and underdamped regimes. We present analytic expressions for the noise-induced coherences in each of these limits, and find that the coherences are maximized when the pumping rates are equal to each other. Taken together, these results  completely characterize the dynamical regimes of the V-system  weakly driven by suddenly turned on incoherent light. (Issues associated with slow turn-on are under consideration \cite{amr2}.)

We note that this work provides new physical insights into the dynamics of the V-system driven by incoherent radiation, which go well beyond those reported previously (including  Ref. \cite{prl14}). In addition to providing novel analytic expressions for density matrix dynamics valid for arbitrary pumping rates, transition dipole orientations, and level splittings [Eqs.~(10)-(16)],  this Article

\begin{enumerate}

\item Establishes the existence of three important dynamical regimes: underdamped, critical, and overdamped, classifiable by the value of the damping parameter $\zeta$ in Eq.~(8b). 

\item Explores the dynamics of the critically damped  V-system (Figs.~7,~8~and~Sec.~IIIC~\cite{note1}).

\item Establishes the presence of asymptotic quasi-stationary coherence in the asymmetric V-system (Sec. IIIA and Figs. 3E-F).

\item Provides a physical interpretation of the quantum state created by incoherent pumping as a mixture of excited eigenstates and their coherent superposition [Eq. (3)].

\item Demonstrates that the noise-induced coherences are maximized when the pumping rates are equal and the transition dipoles are fully aligned.

\end{enumerate}

This paper is organized as follows. Section II presents the BR equations of motion, outlines some general properties of the system dynamics, and gives our general weak-field solution for the V-type system driven by an incoherent radiation field.
Closed-form analytical solutions for the time dynamics of populations and coherences in the limiting case of highly separated excited states ($\zeta \ll 1$, valid for small molecules) are presented in Sec. IIIA. Section IIIB focuses on the case of $\zeta \gg 1$, which holds for large molecules with closely spaced vibronic energy levels. Section IIIC considers a special intermediate regime where the spacing of the excited states is comparable to their radiative decay widths ($\zeta=1$). The results for varying alignment parameter  $p$ and the pumping rates are presented in these subsections.  Section IV concludes with a brief summary of main results and outlines several directions for future work.

\begin{figure}[t!]
	\centering
	\includegraphics[width=0.45\textwidth, trim = 0 0 0 0]{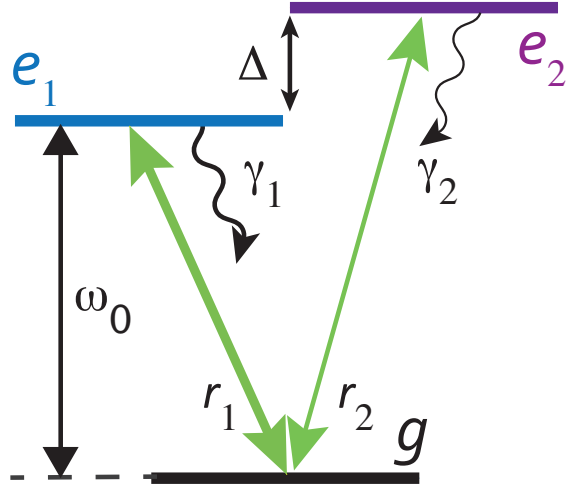}
	\renewcommand{\figurename}{Fig.}
	\caption{Schematic representation of a V-type System. $\Delta$ is the excited state splitting, $\gamma_i$ is the radiative line-width, and $r_i$ is the incoherent pumping rate of excited state $\ket{e_i}$.}
\label{fig:schem}
\end{figure}

\section{Bloch-Redfield Master Equations}
\label{sec:ME}
Consider a V-system interacting with a suddenly turned-on incoherent radiation field, under the Born-Markov approximation.
The dynamics of such a system is governed by the following system of BR master equations \cite{prl14, jcp15}
\begin{subequations}
\beq
\dot{\rho}_{e_i e_i} =-(r_i + \gamma_i)\rho_{e_i e_i} +r_i \rho_{gg}-p(\sqrt{r_1r_2}+\sqrt{\gamma_1\gamma_2})\rho_{e_1e_2}^R
\label{eq:PQME}%
\eeq
\beq
\dot{\rho}_{e_1e_2}=-\frac{1}{2}(r_1+r_2+\gamma_1+\gamma_2)\rho_{e_1e_2}-i\rho_{e_1 e_2}\Delta 
                +\frac{p}{2}\sqrt{r_1 r_2}(2\rho_{gg}-\rho_{e_1
                  e_1}-\rho_{e_2
                  e_2})-\frac{p}{2}\sqrt{\gamma_1\gamma_2}(\rho_{e_1
                  e_1}+\rho_{e_2 e_2})
\label{eq:CQME}%
\eeq
\label{eq:QME}%
\end{subequations}
where here and below atomic ($\hbar=1$) units are used.
In Eq. (\ref{eq:QME}), absorption and stimulated emission processes are parametrized by the incoherent pumping rates of the $\ket{g} \leftrightarrow \ket{e_i}$ transitions $r_i= B_iW(\omega_{g e_i})$, given by the product of the Einstein \emph{B}-coefficients $B_i=\pi |\mu_{g e_i}|^2/(3\epsilon_0)$ and the intensity of the incident blackbody radiation at the corresponding transition  frequencies $W(\omega_{g e_i})$.
Spontaneous emission processes are governed by the radiative decay widths of the excited states, $\gamma_i=\omega_{g e_i}^3|\mu_{g e_i}|^2/(3\pi\epsilon_0c^3)$, $\Delta=\omega_{e_1 e_2}$ gives the excited state energy splitting, and $p=\bm{\mu}_{g e_1}\cdot\bm{\mu}_{g e_2}/(|\mu_{g e_1}| |\mu_{g e_2}|)$ quantifies the alignment of the $\ket{g} \leftrightarrow \ket{e_i}$ transition dipole moments, $\bm{\mu}_{g e_i}$.
The notation $\rho_{e_1e_2}^R$ denotes the real part of $\rho_{e_1e_2}$ while $\rho_{e_1e_2}^I$ denotes the imaginary part.
Here we neglect the environment-induced dephasing and relaxation processes, assuming that the rates of excited state relaxation and dephasing are small compared to those of the radiative processes (absorption, decay and stimulated emission).
Eq.~(\ref{eq:QME}) makes the usual assumptions of weak system-bath coupling and short (delta function) time-correlation in the radiation field \cite{CTbook,SZbook}.


The one-photon coherences between the ground and excited states oscillate extremely quickly and can be decoupled from the dynamics of the rest of the system (the partial secular approximation) \cite{prl14, jcp15}.
The quantum master equations [Eq. \ref{eq:PQME}] for the populations contain secular rate-law terms which depend only on the diagonal density matrix elements, $\rho_{e_1e_1}$, and coherence-dependent non-secular terms (proportional to $\rho_{e_1e_2}$) that lead to the breaking of detailed balance responsible for the proposed efficiency enhancement in quantum heat engines \cite{Scully11, Scully13}.
The rate-law terms contain the effects of the independent $\ket{g} \leftrightarrow \ket{e_i}$ transitions while the coherence-dependent terms show the effect of interference between the two pathways on the emission processes.
The dynamics of the  V-system can be understood in all regimes in terms of the interplay of interference and rate-law transitions.

The quantum master equations can be recast in the Liouville representation in terms of the state vector $\mathbf{x}=[\rho_{e_1e_1},\rho_{e_2e_2},\rho_{e_1e_2}^R,\rho_{e_1e_2}^I]^T$ where $\rho_{e_1e_2}^R$ and  $\rho_{e_1e_2}^I$ are the real and imaginary parts of the excited state coherence, respectively.
Substituting the normalization condition ($\rho_{gg}=1-\rho_{e_1e_1}-\rho_{e_2e_2}$) into Eq. (\ref{eq:QME}) allows us to write the master equations in the following inhomogeneous form:
\begin{subequations}
\beq
\frac{d}{dt} \mathbf{x} = A\mathbf{x}+ \mathbf{d}
\label{eq:MLRME}%
\eeq
\beq
A= \left( 
\begin{array}{cccc}
-(r_1+\gamma_1) & -r_1 & -p\left(\sqrt{r_1r_2}+\sqrt{\gamma_1\gamma_2}\right) & 0 \\
-r_2 & -(r_2+\gamma_2) & -p\left(\sqrt{r_1r_2}+\sqrt{\gamma_1\gamma_2}\right) & 0 \\
-\frac{p}{2}\left(3\sqrt{r_1r_2}+\sqrt{\gamma_1\gamma_2}\right) & -\frac{p}{2}\left(3\sqrt{r_1r_2}+\sqrt{\gamma_1\gamma_2}\right) & -(\rbar +\gbar) & \Delta \\
0 & 0 & -\Delta & -(\rbar +\gbar)
\end{array}\right)
\label{eq:Amat}%
\eeq
\beq
\mathbf{d}=\left(\begin{array}{c}
r_1 \\ r_2 \\ p\sqrt{r_1r_2}\\0
\end{array} \right)
\label{eq:dvec}%
\eeq
\label{eq:VME}%
\end{subequations}
where $\rbar= \frac{1}{2}(r_1+r_2)$ and $\gbar = \frac{1}{2}(\gamma_1+\gamma_2)$ are the mean pumping rate and radiative decay width, respectively.
In Eq. (\ref{eq:VME}), $A$ is a coefficient matrix, analogous to the Liouville superoperator, and $\mathbf{d}$ is a driving vector containing the contributions of radiative excitation from the ground state $\ket{g}$.
The driving vector, $\mathbf{d}$ (Eq. \ref{eq:dvec}), indicates that the incoherent radiation drives the V-system into the following mixed state on the excited state manifold:
\beq
\rho_d \propto (1-p)(r_1\ket{e_1}\bra{e_1}+r_2\ket{e_1}\bra{e_2})+p\ket{\phi_+}\bra{\phi_+}
\label{eq:rhod}
\eeq
where $\ket{\phi_+}=\frac{1}{\sqrt{2\rbar}}(\sqrt{r_1}\ket{e_1}+\sqrt{r_2}\ket{e_2})$ is an in-phase coherent superposition of the energy eigenstates.
Equivalently, only the real part of the coherences, $\rho_{e_1e_2}^R$, is directly excited from the ground state [see Eq. (\ref{eq:CQME})].
Equation (\ref{eq:rhod}) illustrates the important role of the alignment factor $p$ in parametrizing the preparation of the coherent superposition $\ket{\phi_+}$.
In contrast, the Pauli rate-law equations predict excitation into the incoherent mixture of excited states:
\beq
\rho_\text{incoh} \propto r_1\ket{e_1}\bra{e_1}+r_2\ket{e_2}\bra{e_2}
\label{eq:rhoPauli}
\eeq
that equilibrates to the mixture:
\beq
\rho_\text{eq} \propto \ket{e_1}\bra{e_1}+\ket{e_2}\bra{e_2}
\label{eq:rhoeq}
\eeq
Physically, $\ket{\phi_+}$ is generated by a simultaneous excitation from the ground state to the two excited states  $\ket{e_1}$ and $\ket{e_2}$.
In the limit of orthogonal transition dipole moments ($p = 0$), Eq. (\ref{eq:rhod}) gives an incoherent mixture of the energy eigenstates [Eq. (\ref{eq:rhoPauli})] while in the opposite limit of parallel dipole moments ($p= 1$), the pure in-phase coherent superposition, $\ket{\phi_+}\bra{\phi_+}$, is produced.

Using the standard variation of parameters procedure \cite{BDP}, the solutions to the master equations (\eqs{eq:VME}) with initial conditions corresponding to excitation from the ground state ($\mathbf{x}(t=0)=\mathbf{0}$) is given by
\beq
\mathbf{x}(t) = \int^t_0 ds e^{A(t-s)}\mathbf{d} \to \sum_{i=1}^4 \int^t_0 ds (\mathbf{v}_i\cdot\mathbf{d})e^{\lambda_i (t-s)} \mathbf{v}_i
\label{eq:GenSol}
\eeq
where $\lambda_i$ is the $i^{th}$ eigenvalue of $A$ and $\mathbf{v}_i$ the corresponding eigenvector.
Eq. \ref{eq:GenSol} relates the eigenvalues, $\{\lambda_i\}$, of $A$ to the timescales of the system's evolution $\tau_i= \Real(\lambda_i)^{-1}$ and the frequencies of its oscillations $\omega_i =\Imag(\lambda_i)$.

In general, the spectrum of the coefficient matrix, A, (Eq. \ref{eq:Amat}) is complicated.
However, we note that the intensity of the incident blackbody radiation encountered in nature (represented here by the effective thermal occupation number $\nbar=\frac{r_i}{\gamma_i}$), is typically very weak, $\nbar \ll 1$ (the weak pumping limit).
This allows us to neglect the terms of order $r_i$ in Eq. (\ref{eq:Amat}) and greatly simplifies the calculation of the spectrum, $\{ \lambda_i \}$ of $A$, to give:
\begin{equation}
	\lambda_i = -\bar{\gamma}\pm \sqrt{\frac{(\bar{\gamma}^2-\Delta_p^2)\pm\sqrt{(\bar{\gamma}^2-\Delta_p^2)^2+\Delta^2 (\gamma_1-\gamma_2)^2}}{2}}
	\label{eq:lambda}
\end{equation}
where $\Delta_p =\sqrt{\Delta^2+(1-p^2)\gamma_1\gamma_2}$ and $\gbar=\frac{1}{2}(\gamma_1+\gamma_2)$.
The algebraic details of the derivation of Eq. (\ref{eq:lambda}) are given in Eq. (\ref{app:spec}) of the Appendix.

The evolution of the coherences, $\rho_{e_1e_2}$, can be understood in analogy to the behavior of a damped harmonic oscillator.
To illustrate this analogy, consider a system with symmetrically broadened excited states ($\gamma_1=\gamma_2$).
In this case, Eq. (\ref{eq:lambda}) gives one degenerate negative root ($\lambda_1=-\gbar$) and a pair of (possibly complex) roots ($\lambda_{2,3}=-\gbar\pm\sqrt{(\gbar^2-\Delta_p^2)}$) analogous to a damped oscillator with natural frequency $\Delta_p$ and damping constant $\gbar$ \cite{Morin}.

Motivated by the similarities to the damped oscillator, we rewrite Eq. (\ref{eq:lambda}) as
\beqs
\beq
\lambda_i= -\gbar \pm \Delta_p \sqrt{\zeta^2-1} \sqrt{\frac{1 \pm \sqrt{1+\eta^2}}{2}}
\label{eq:lamDO}%
\eeq
\beq
\zeta =\frac{\gbar}{\Delta_p}
\label{eq:zeta}%
\eeq
\beq
\eta =\frac{\Delta|\gamma_1-\gamma_2|}{|\gbar^2-\Delta_p^2|},
\label{eq:eta}%
\eeq
\eeqs
where the parameter $\zeta$ plays the same role as the damping coefficient of a classical damped oscillator.
On the other hand, Eq. (\ref{eq:eta}) defines a second parameter, $\eta$, which quantifies the deviation from the damped oscillator spectrum, (with $\eta=0$ corresponding to the symmetrically pumped V-system, which is equivalent to the damped harmonic oscillator).
As such, Eq. (\cref{eq:eta}) relates the asymmetry of the excited states (in terms of their splitting and radiative broadening) to their anharmonicity, $\eta$.

To clarify the effects of $\eta$ on the spectrum Eq. (\cref{eq:lamDO}) and hence its deviation from that of a harmonic oscillator consider the positive and negative branches of the $\eta$ dependent term ($\sqrt{\frac{1\pm\sqrt{1+\eta^2}}{2}}$) separately.
In the underdamped regime, the negative branch in Eq. (\ref{eq:lamDO}) splits the degenerate eigenvalue to give two new relaxation rates, one slower than the original rate and the other faster, while the positive branch increases the frequency of the system's oscillations.
On the other hand, the negative branch ($1-\sqrt{1+\eta^2}$) leads to asymmetry induced oscillations in the overdamped regime ($\zeta \gg 1$).
The positive branch modifies the relaxation rates in this regime, increasing the faster rate while slowing down the slower rate.

To separate the effects of the asymmetry of $\gamma_1$ and $\gamma_2$ from their magnitude, $2\gbar$ we introduce the parameter $\beta$ such that 
\beqs
\beq
\gamma_1 =  2\gbar\sin^2(\beta)%
\eeq
\beq
\gamma_2=2\gbar\cos^2(\beta)%
\eeq
\beq
\eta = 2|\cos(2\beta)|\frac{\Delta}{\Delta_p}\frac{\zeta}{|\zeta^2-1|}%
\eeq
\label{eq:betas}%
\eeqs
For fixed $p$, $\beta$ provides a convenient independent measure of the asymmetry of the transitions, with decreasing $\beta \in [0,\frac{\pi}{4}]$ indicating a larger difference in decay rates $|\gamma_1-\gamma_2|$.

\begin{figure}[ht]
	\centering
	\includegraphics[width=\textwidth, trim = 0 0 0 0]{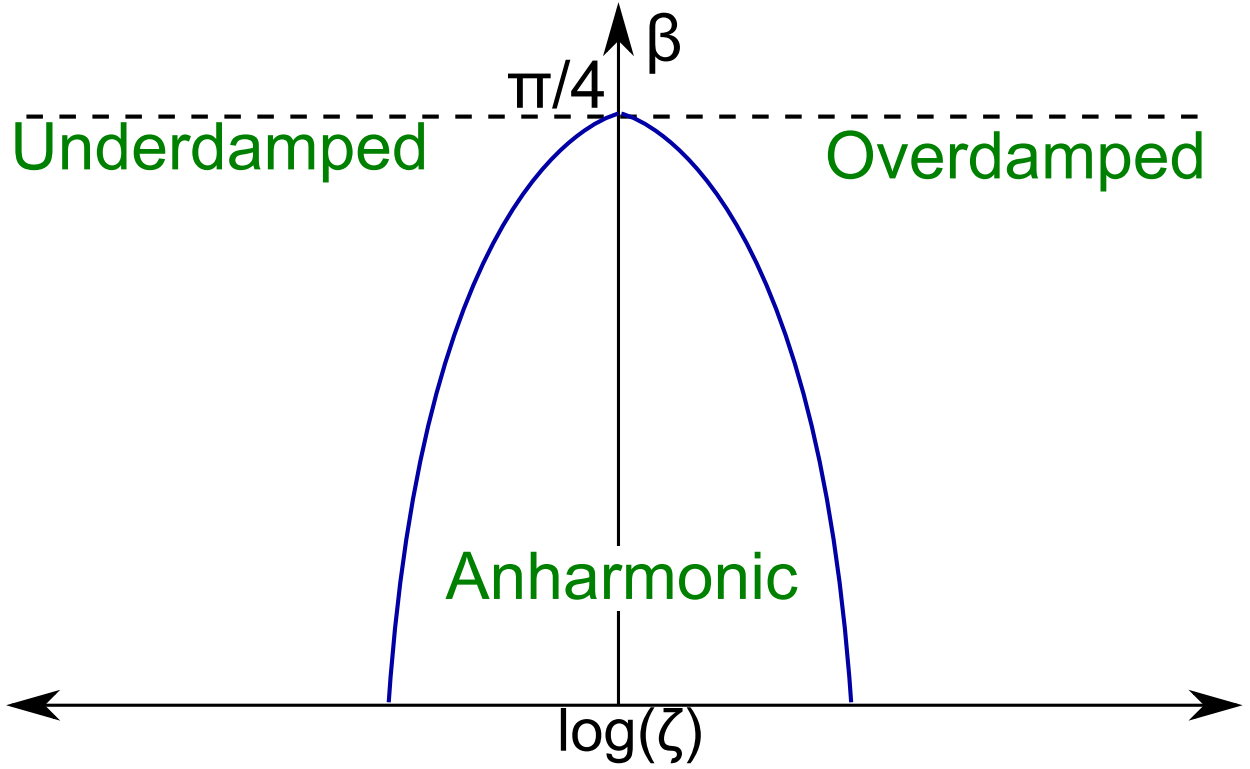}
	\renewcommand{\figurename}{Fig.}
	\caption{A "phase diagram" depicting the regions in parameter space corresponding to the regimes of the V-system. The underdamped regime $\zeta <1$ shows damped oscillatory coherences, while the overdamped and critically damped regimes $\zeta \geq1$ show no oscillations. As in Eq. (\ref{eq:zeta}), $\zeta =\frac{\gbar}{\Delta_p}$ and $\beta$ is defined by $\gamma_1=2\gbar\cos^2(\beta)$, $\gamma_1=2\gbar\sin^2(\beta)$ (Eq. \ref{eq:betas}). Note that the difference in line widths ($|\gamma_1-\gamma_2|$) increases as $\beta \in[0, \frac{\pi}{4}]$ decreases.}
\label{fig:Phase}
\end{figure}

In previously reported analytical results on the V-system, derived for symmetrically broadened states ($\gamma_1=\gamma=\gamma_2$) and collinear dipole moments ($p=1$), the parameter ${\Delta}/{\gamma}$ was introduced \cite{prl14}, analogous to $\zeta^{-1}={\Delta_p}/{\gbar}$. 
However, to highlight the similarities to the damped oscillator we opt for the parameter $\zeta$.  Figure \ref{fig:Phase} sketches the various dynamical regimes of the V-system excited by incoherent light. The vertical line corresponds to $\zeta=1$ and separates the underdamped and underdamped regimes. The region below the horizontal line $\beta=\pi/4$ corresponds to the asymmetric V-system, with the degree of anharmonicity increasing as $\beta$ decreases. As discussed below,  the coherent dynamics in the strongly anharmonic region (corresponding to $\beta\to 0$) displays asymmetry-induced quasistationary coherences that are not present in the symmetric V-system.


\section{Analytical Solutions}
\label{sec:Analyt}

\subsection{Underdamped Regime $\zeta \ll 1$}
\label{subsec:Under}

Consider first a weakly damped system with $\zeta \ll 1$ (or equivalently $\frac{\Delta_p}{\gbar} \gg1$).
Physically, this corresponds to a system with a large excited state splitting, $\Delta$, such as a small molecule with few vibronic degrees of freedom.
In this limit, the spectrum, $\{\lambda_i \}$, and corresponding normal modes, $\{\mathbf{v}_i\}$, of $A$ can be calculated through a binomial expansion of Eq. (\ref{eq:lamDO}) and  substituted into Eq. (\ref{eq:GenSol}) to give the density matrix elements.
(The details of this calculation are provided in section Eq. (\ref{app:UDamp}) of the Appendix.)
This gives
\beqs 
\beq
\rho_{e_i,e_i}(t)=\nbar_e (1-e^{-\gamma_it})
\label{eq:UDP}%
\eeq
\beq
\rho_{e_1,e_2}^R(t)=\frac{p \sqrt{r_1r_2}}{\Delta_p}e^{-\gbar t}\sin(\Delta_p t)
\label{eq:UDCR}%
\eeq
\beq
\rho_{e_1,e_2}^I(t)=\frac{p \sqrt{r_1r_2}}{\Delta_p}\left( e^{-\gbar t}(\cos(\Delta_p t)-1)-\frac{e^{-\gamma_1 t}-e^{-\gamma_2t}}{2} \right)
\label{eq:UDCI}%
\eeq
\label{eq:UDamp}%
\eeqs
where $\nbar_e$ is the effective thermal occupation ({\it e.g.} $\nbar_e=\alpha[e^{\frac{\omega_0}{k_BT}}-1]^{-1}=\alpha\nbar$. Here $\alpha$ is some scalar and $\nbar$ is the thermal occupation of  electromagnetic field modes at the ground-to-excited manifold transition frequency, $\omega_0$ (see Fig. 1). We have assumed, without loss of generality, that $\gamma_1 < \gamma_2$.
We use the same occupation number for the $\ket{g} \leftrightarrow \ket{e_1}$ and $\ket{g} \leftrightarrow \ket{e_2}$ transitions.
This is justified since $\Delta \ll \omega_0$ under the Wigner-Weisskopf approximation \cite{prl14, SZbook}, leading to $r_i=\nbar_e \gamma_i$ or $r_1/\gamma_1 = r_2/\gamma_2=\nbar_e$, which guarantees that a coherence-free (canonical-like) state of the V-system is reached in the limit $t\to \infty$ for non-degenerate excited states \cite{Agarwal01}.

\begin{figure}[ht]
	\centering
	\includegraphics[width=\textwidth, trim = 0 0 0 0]{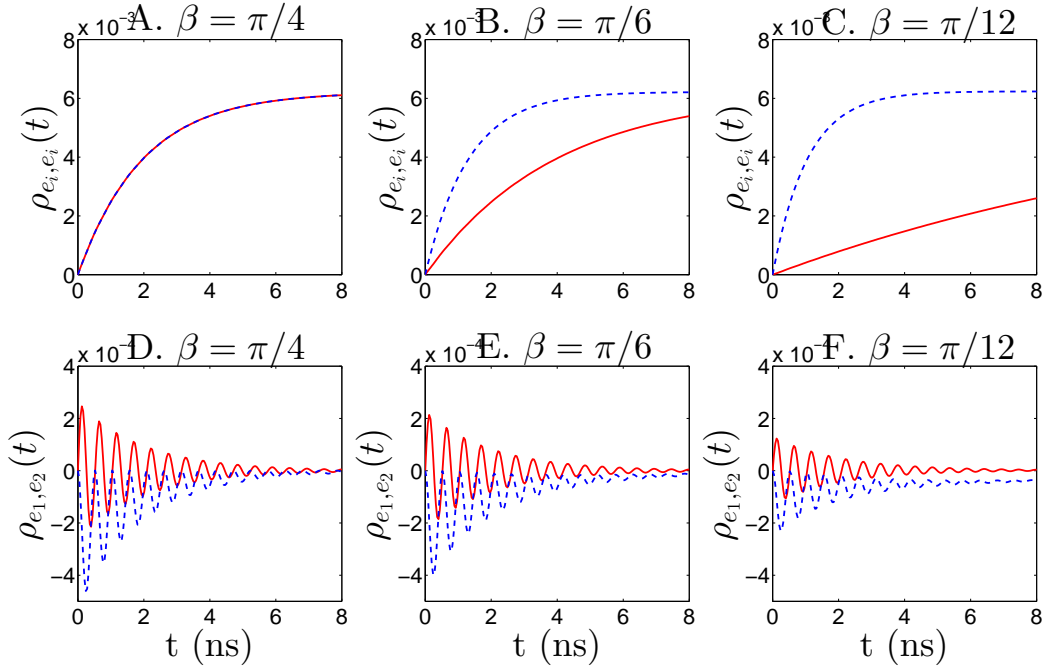}
	\renewcommand{\figurename}{Fig.}
	\caption{Evolution of populations and coherences of an underdamped V-system ($\zeta \ll 1$), typical of  small molecule, evaluated with aligned transition dipole moments ($p=1$). Panels A-C show the evolution of the populations $\rho_{e_1e_1}$ (solid) and $\rho_{e_2e_2}$ (dashed) where state $\ket{e_2}$ has the highest energy. Subplots D-F show the evolution of the real ($\rho_{e_1,e_2}^R$ solid line) and imaginary ($\rho_{e_1,e_2}^I$ dashed line) coherences. Here $\gbar=1.0$ and $\Delta=12.0$, with varying pumping rate asymmetry $\gamma_1=\gbar \sin^2(\beta)$ and $\gamma_2=\gbar \cos^2(\beta)$.}
\label{fig:1}
\end{figure}

The evolution of the density matrix elements is illustrated in Fig. \ref{fig:1}.
In this underdamped limit, Eq. (\ref{eq:UDP}) and Fig. \ref{fig:1}A-C show the evolution of the excited state populations, $\rho_{e_ie_i}$, to a steady state value, at a rate $\gamma_i$, as predicted by the Pauli rate equations.
The lack of $p$ dependence in Eq. (\ref{eq:UDP}) and the adherence to the dynamics predicted by the secular rate-law equations indicates that the populations of a V-type system in the underdamped regime is unaffected by the coherences.
In other words, the time dynamics of the diagonal elements of the density matrix is described equally well by the Pauli-type rate-law equations [i.e. Eq. (\ref{eq:QME}) with $p=0$].

By contrast, \eqs{eq:UDCR} and Figs. \ref{fig:1}D-F show damped oscillatory coherences, $\rho_{e_1e_2}$.
 Equations (\ref{eq:UDCR}) and (\ref{eq:UDCI}) can be understood in terms of the coherent $\ket{\phi_+} = \frac{1}{\sqrt{2\rbar}}(r_1 \ket{e_1} +r_2 \ket{e_2})$ component of the mixture, $\rho_d$, prepared by the incoherent radiation, Eq. (\ref{eq:rhod}).
The $\ket{\phi_+}$ component of $\rho_d$ is initially in an in-phase (positive) superposition of the excited states.
The unitary evolution induced by the (isolated) V-type system Hamiltonian causes the components in $\ket{e_1}$ and $\ket{e_2}$ to accrue a relative phase with frequency $\Delta$, producing the oscillations seen in Eqs. (\ref{eq:UDCR}) and (\ref{eq:UDCI}) and in Figs. \ref{fig:1}D-F.
However, the interaction with the radiation field leads to a decay of the coherent wavepacket through spontaneous and stimulated emission, leading to exponential decay at a rate $\gbar$, as seen in Eqs. (\ref{eq:UDCR}) and (\ref{eq:UDCI}) and Figs. \ref{fig:1}.D-F.
Eventually, the interaction with the radiation field destroys the coherent  $\ket{\phi_+}$ component of the mixture generated by the blackbody radiation $\rho_d$ Eq. (\ref{eq:rhod}) to give the incoherent mixture predicted by the rate-law equations, $\rho_{eq}$ [Eq. (\ref{eq:rhoeq}) or (\ref{eq:QME}) under the secular approximation ($p=0$)].

A unique feature of the asymmetric V-system  ($\beta \neq \pi/4$) is the presence of asymptotic quasistationary imaginary coherences (Figs. \ref{fig:1}E-F).
These are contained in the non-oscillatory exponentials in Eq. (\ref{eq:UDCI}) and are a manifestation of the slow equilibration of one of the states.
When one state, say $\ket{e_2}$, equilibrates much more quickly than the other (i.e. $\gamma_2 \gg \gamma_1$) then the large population of $\ket{e_2}$, $\rho_{e_2e_2}$, suppresses the real part of the coherences, $\rho_{e_1e_2}^R$, through the non-secular term coupling $\rho_{e_1e_2}^R$ to $\rho_{e_2e_2}$ in Eq. (\ref{eq:CQME}).
Physically, this corresponds to constructive interference in the decay processes of in phase superpositions (e.g. $\ket{\phi_+}$), which are characterized by  $\rho_{e_1e_2}^R >0$.
Consequently, the real coherences decay more quickly than the imaginary coherences which survive for longer times due to the slow equilibration of $\ket{e_1}$.

Since the coherences are only present in the $\ket{\phi_+}$ components of $\rho_d$, the $p$ dependence of the amplitude of coherences in Eqs. (\ref{eq:UDCR}) and (\ref{eq:UDCI}) reflects the amount of $\ket{\phi_+}$ in the $\rho_d$ mixture, Eq. (\ref{eq:rhod}).
Further, the magnitude of coherences in a pure $\ket{\phi_+}=\frac{1}{\sqrt{2\rbar}}(r_1\ket{e_1}+r_2\ket{e_2})$ state is proportional to $\sqrt{r_1r_2}$.
This produces the $\sqrt{r_1r_2}$ scaling in the amplitude of the coherences, $\rho_{e_1e_2}$ in Eqs. (\ref{eq:UDCR}) and (\ref{eq:UDCI}) and hence the decreased amplitude of coherences as the difference between $\gamma_1 = \frac{r_1}{\nbar}$ and $\gamma_2=\frac{r_2}{\nbar}$ increases in Figs. \ref{fig:1}.D-F.

\begin{figure}[ht]
	\centering
	\includegraphics[width=\textwidth, trim = 0 0 0 0]{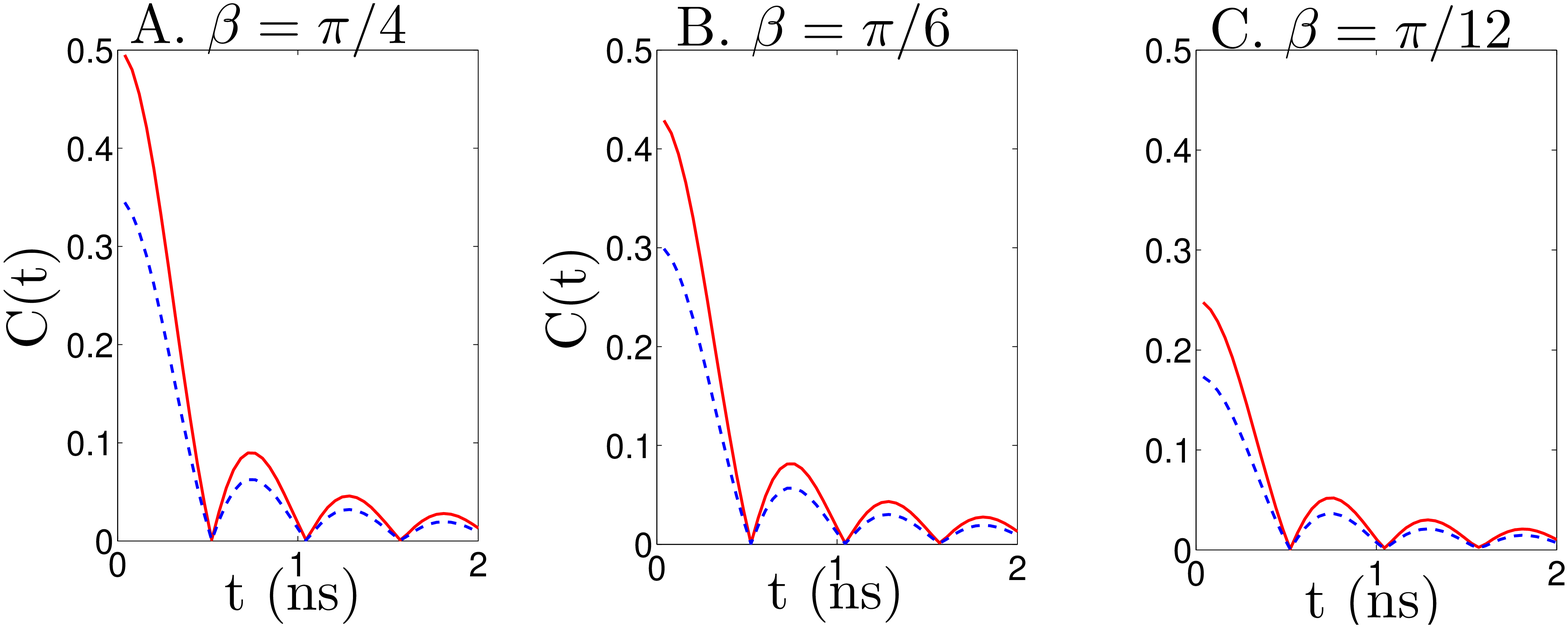}
	\renewcommand{\figurename}{Fig.}
	\caption{Evolution of the coherence ratio $C(t)=\frac{|\rho_{e_1,e_2}|}{\rho_{e_1,e_1}+\rho_{e_2,e_2}}$ of an underdamped V-system ($\zeta \ll 1$). Here $\gbar=1.0$ and $\Delta=12.0$, with varying pumping rate asymmetry $\gamma_1=\gbar \sin^2(\beta)$ and $\gamma_2=\gbar \cos^2(\beta)$ and dipole alignment, $p=1.0$ (solid) and $p=0.7$ (dashed). }
\label{fig:2}
\end{figure}

The coherence ratio $C(t)=\frac{|\rho_{e_1,e_2}|}{\rho_{e_1,e_1}+\rho_{e_2,e_2}}$ provides a convenient method of quantifying the magnitude of the coherences relative to the populations of the corresponding states \cite{prl14,Zaheen,pra14}.
Neglecting the weak asymptotic imaginary coherences that appear in highly asymmetric systems, the coherence ratio for an underdamped V-system is given by
\beq
C(t)\approx \left(\frac{p\sqrt{\gamma_1\gamma_2}}{\Delta_p}\right)\left(\frac{e^{-\gbar t}}{1-e^{-\gbar t}}\right)\sin^2\left(\frac{\Delta_p t}{2} \right)
\label{eq:UDC}
\eeq
and is shown in Fig. \ref{fig:2}. The $C$-ratios exhibit damped oscillations, decreasing at short time as $1/(\Delta_p t)$, extending our earlier result \cite{pra14,pra14,Zaheen}.

 It follows from \cref{eq:UDC} that $C(t)$ is maximized when the decay widths, $\gamma_i$, (and hence the pumping rates $r_i$) are equal.
Furthermore, it can be seen that the amplitude of the coherence ratio scales linearly with the alignment parameter, $p$, and inversely with excited state splitting, $\Delta$ contained in $\Delta_p$.
This scaling presented in the prefactor of \cref{eq:UDC}, can also be seen in \cref{fig:2}.
Furthermore, the effects of the asymmetry in $\gamma_i$, that are not accounted for in \cref{eq:UDC}, manifest in Fig. \ref{fig:2}C in the small residual value of $C(t)$ between $\tau_1=\gamma_1^{-1}$ and $\tau_2=\gamma_2^{-1}$.
 
\subsection{Overdamped Regime $\zeta \gg 1$}
\label{subsec:Over}

\begin{figure}[ht]
	\centering
	\includegraphics[width=\textwidth, trim = 0 0 0 0]{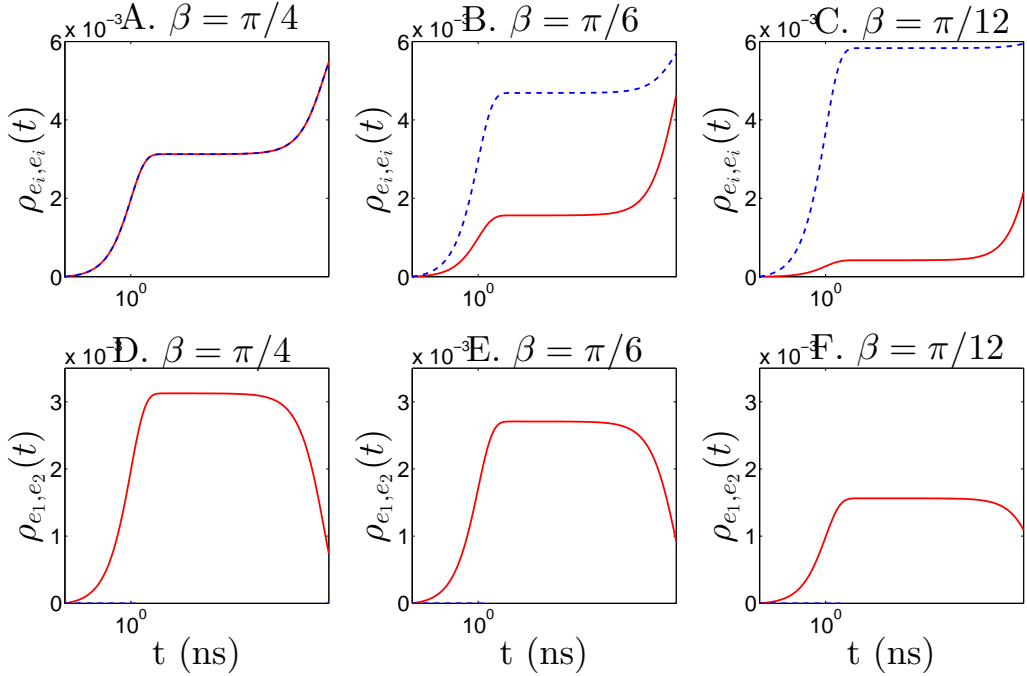}
	\renewcommand{\figurename}{Fig.}
	\caption{Evolution of populations and coherences of an overdamped V-system ($\zeta \gg 1$) evaluated with aligned transition dipole moment operators ($p=1$).Panels A-C show the evolution of the populations $\rho_{e_1e_1}$ (solid) and $\rho_{e_2e_2}$ (dashed) where state $\ket{e_2}$ has the highest energy. Panels D-F show the evolution of the real ($\rho_{e_1,e_2}^R$, solid) and imaginary ($\rho_{e_1,e_2}^I$, dashed) coherences. Note that the imaginary coherences are heavily suppressed and nearly vanish throughout the system's evolution. Here $\gbar=0.5$, and $\Delta=0.0012$, with varying pumping-rate asymmetry $\gamma_1=\gbar \sin^2(\beta)$ and $\gamma_2=\gbar \cos^2(\beta)$}
\label{fig:3}
\end{figure}

Consider now the case of a system that is strongly damped ($\zeta \gg 1$ or equivalently $\frac{\Delta_p}{\gbar} \ll 1$). Using a similar Taylor expansion of \cref{eq:lamDO} as in the underdamped case, the evolution of the density matrix is obtained after some algebra, detailed in section \cref{app:ODamp} of the Appendix, as
\beqs
\beq
\rho_{e_i,e_i}(t)=\frac{1}{2\gbar}\left[r_i(1-e^{-2\gbar t})+r_{j}(1-e^{-\frac{\Delta_p^2}{2\gbar}t}) \right]
\label{eq:ODP}%
\eeq
\beq
\rho_{e_1,e_2}^R(t)=\frac{p \sqrt{r_1r_2}}{2\gbar}\left[e^{-\frac{\Delta_p^2}{2\gbar}t}-e^{-2\gbar t}\right]
\label{eq:ODCR}%
\eeq
\label{eq:ODamp}%
\eeqs
where $i,j=1,2$ and $i\neq j$.
The imaginary coherence, $\rho_{e_1,e_2}^I=\Im{\rho_{e_1e_2}}$, is suppressed by a factor $\zeta^{-1} \ll 1$.
These density matrix elements are plotted in Fig. \ref{fig:3}.

\Cref{eq:ODamp} and \cref{fig:3} show that, in the overdamped regime, the V-system first evolves to the mixture, $\rho_d$ [\cref{eq:rhod}], excited by the incoherent radiation over a time $\tau_{short}=(2\gbar)^{-1}$.
Then $\rho_d$ decays to the incoherent mixture, $\rho_{eq}$ [\cref{eq:rhoeq}], generated by the rate-law model.
However, \cref{eq:ODamp} and \cref{fig:3} show that the coherences (and hence the coherent $\ket{\phi_+}$ component of $\rho_d$) are remarkably long-lived, surviving for
\beq
\tau_{long}=\frac{2\gbar}{\Delta_p^2}
\label{eq:taulong}
\eeq
In particular, \cref{eq:taulong} indicates that $\tau_{long} \to \infty$ as $\Delta_p \to 0$ and $\gbar \neq 0$.
In the $\Delta_p=0$ case, $\rho_d$ is dynamically stable and survives in the long time steady state \cite{Scully06}. Noting that $\Delta_p \to 0$ indicates that $\Delta=0$ and $p=1$, this is equivalent to the steady state coherences observed in previous investigations of the V-system \cite{prl14, Scully06}, which give rise to coherent population trapping and multiple (initial-state-dependent) steady states \cite{Scully06,Kurizki}. 

In a degenerate V-system (i.e., $\Delta=0$), the coherent superposition $\ket{\phi_+}$ accrues no relative phase, corresponding to the vanishing of the coupling between the real and imaginary coherences in \cref{eq:CQME}.
As $\Delta \to 0$, the evolution of the density matrix is determined by the interaction between the components of $\rho_d$.
In particular, the eigenstate  components, $\ket{e_1}$ and $\ket{e_2}$, suppress the coherent $\ket{\phi_+}$ component through the population-coherence coupling term in \cref{eq:CQME}.
If $p \neq 1$ then \cref{eq:rhod} indicates that $\ket{e_1}$ and $\ket{e_2}$ will be present in $\rho_d$, increasing the decoherence rate.
However, if $p=1$ and $\Delta=0$ then the coherent pure state $\rho_d =\ket{\phi_+}\bra{\phi_+}$ does not decay, producing the steady state coherences observed in \cref{eq:ODamp} \cite{Scully06}.

\begin{figure}[ht]
	\centering
	\includegraphics[width=\textwidth, trim = 0 0 0 0]{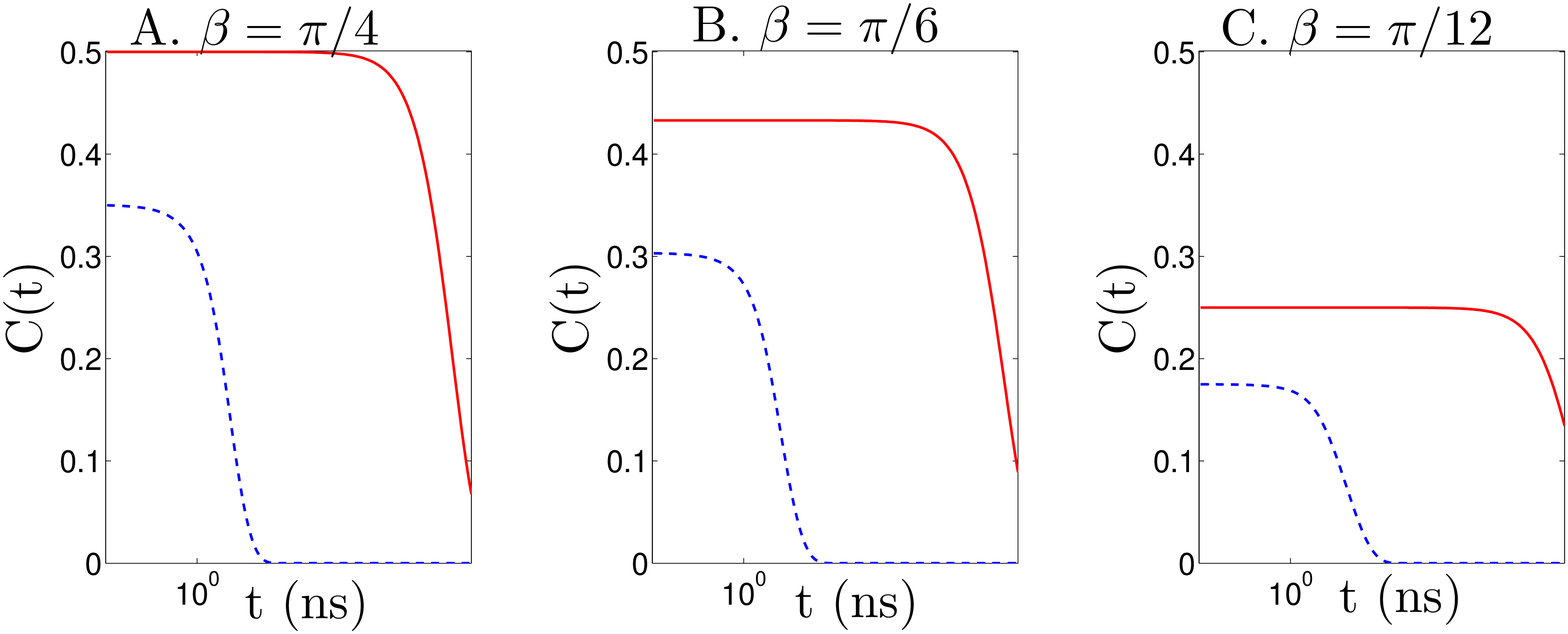}
	\renewcommand{\figurename}{Fig.}
	\caption{Evolution of the coherence ratio $C(t)=\frac{|\rho_{e_1,e_2}|}{\rho_{e_1,e_1}+\rho_{e_2,e_2}}$ of an overdamped V-system ($\zeta \gg 1$). Here $\gbar=1.0$, and $\Delta=0.0012$, with varying pumping-rate asymmetry $\gamma_1=\gbar \sin^2(\beta)$ and $\gamma_2=\gbar \cos^2(\beta)$ and dipole alignment, $p=1.0$ (solid) and $p=0.7$ (dashed). }
\label{fig:4}
\end{figure}

The coherence ratio, $C(t)=\frac{|\rho_{e_1e_2}|}{\rho_{e_1e_1}+\rho_{e_2e_2}}$, of a V-system in the overdamped regime is given by
\beq
C(t)=\frac{p\sqrt{r_1r_2}}{2\rbar}\left(\frac{e^{-\frac{\Delta_p^2}{2\gbar}t}-e^{-2\gbar t}}{2-e^{-\frac{\Delta_p^2}{2\gbar}t}-e^{-2\gbar t}}\right)
\label{eq:ODC}
\eeq
A plot of $C(t)$ is shown in \cref{fig:4}.
Both \Cref{eq:ODC,fig:4} illustrate a notably long-lived substantial $C(t)$ followed by the collapse of  $\rho_d$ to the incoherent mixture $\rho_{eq}$ in time $\tau_{long}$.
Furthermore, note that \cref{eq:UDC,eq:ODC} both show the same scaling dependence on $p\sqrt{r_1r_2}$.
That is, as in the underdamped regime, the $p$ dependence reflects the dependence on the amount of $\ket{\phi_+}$ present in $\rho_d$ while $\sqrt{r_1r_2}$ gives the coherence of the $\ket{\phi_+}\bra{\phi_+}$ pure state.

\subsection{Critical Regime $\zeta = 1$, $\eta=0$}
\label{subsec:Crit}

\begin{figure}[ht]
	\centering
	\includegraphics[width=\textwidth, trim = 0 0 0 0]{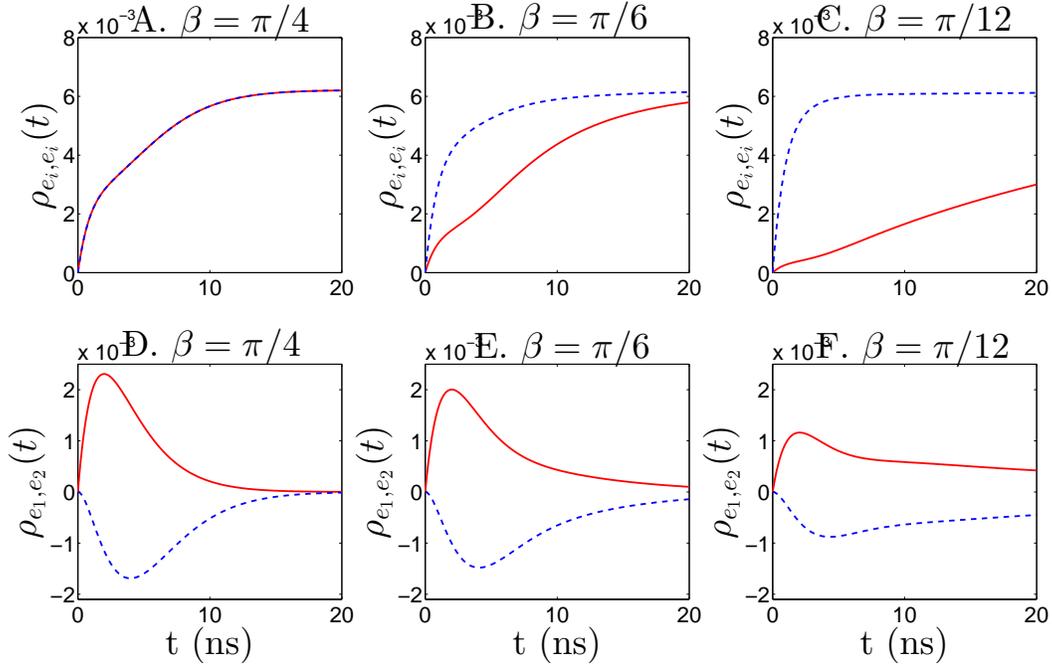}
	\renewcommand{\figurename}{Fig.}
	\caption{Evolution of populations and coherences of a critically damped V-system ($\zeta \gg 1$) evaluated with aligned transition dipole moment operators ($p=1$). Panels A-C show the evolution of the populations $\rho_{e_1e_1}$ (solid) and $\rho_{e_2e_2}$ (dashed) where state $\ket{e_2}$ has the highest energy. Panels D-F show the evolution of the real ($\rho_{e_1,e_2}^R$, solid) and imaginary ($\rho_{e_1,e_2}^I$, dashed) coherences. Here $\gbar=1.0$, and $\Delta=1.0$, with varying pumping-rate asymmetry $\gamma_1=\gbar \sin^2(\beta)$ and $\gamma_2=\gbar \cos^2(\beta)$.}
\label{fig:5}
\end{figure}

Finally, to connect the overdamped and underdamped regions, consider a special case between them.
In particular, when $\zeta=\frac{\gbar}{\Delta_p}=1$ (or equivalently $\frac{\Delta_p}{\gbar}$) and $\eta=\frac{\Delta|\gamma_1-\gamma_2|}{|\gbar^2-\Delta_p^2|}$ the spectrum, $\{ \lambda_i\}$, of $A$ (\cref{eq:lamDO}) becomes completely degenerate with one eigenvalue $\gbar$.
These constraints are equivalent to $\gamma_1=\gamma=\gamma_2$ and $\Delta=p\gamma$.
After some algebra, detailed in section \cref{app:CDamp} of the appendix, the matrix $A$ can be put into Jordan Canonical Form and substituted into \cref{eq:GenSol} to give the density matrix elements.
\beqs
\beq
\rho_{e_i,e_i}(t)=\nbar_e (1-e^{-\gamma t}+\frac{p^2\gamma^2}{2}t^2e^{-\gamma t})
\label{eq:CDP}%
\eeq
\beq
\rho_{e_1,e_2}^R(t)=\nbar_e p\gamma t e^{-\gamma t}
\label{eq:CDCR}%
\eeq
\beq
\rho_{e_1,e_2}^I(t)=-\nbar_e \frac{p^2\gamma^2}{2}t^2 e^{-\gamma t}
\label{eq:CDCI}%
\eeq
\label{eq:CDamp}%
\eeqs
%
The long-time ($t \to \infty$) limit of \cref{eq:CDamp} yields the incoherent mixture of excited states predicted by the Pauli rate equations, $\rho_{eq}$ (\cref{eq:rhoeq}).
However, in contrast to the underdamped regime, the coherences alter the approach of the system to its equilibrated state.
In \cref{eq:CDamp}, this can be seen in the terms proportional to $t$ and $t^2$.
The altered dynamics of the density matrix elements are also clearly seen in \cref{fig:5}.
In particular, the slowly equilibrating populations, $\rho_{e_1e_1}$, show a marked "bend" in their approach to the steady state, resulting in a transient suppression of population compared to their underdamped or coherence free evolution.
More generally, as the damping of the system becomes stronger, the effects of the coherences become more prominent and long-lasting.
In the critically damped regime, no quasistationary behavior is observed.

\begin{figure}[ht]
	\centering
	\includegraphics[width=\textwidth, trim = 0 0 0 0]{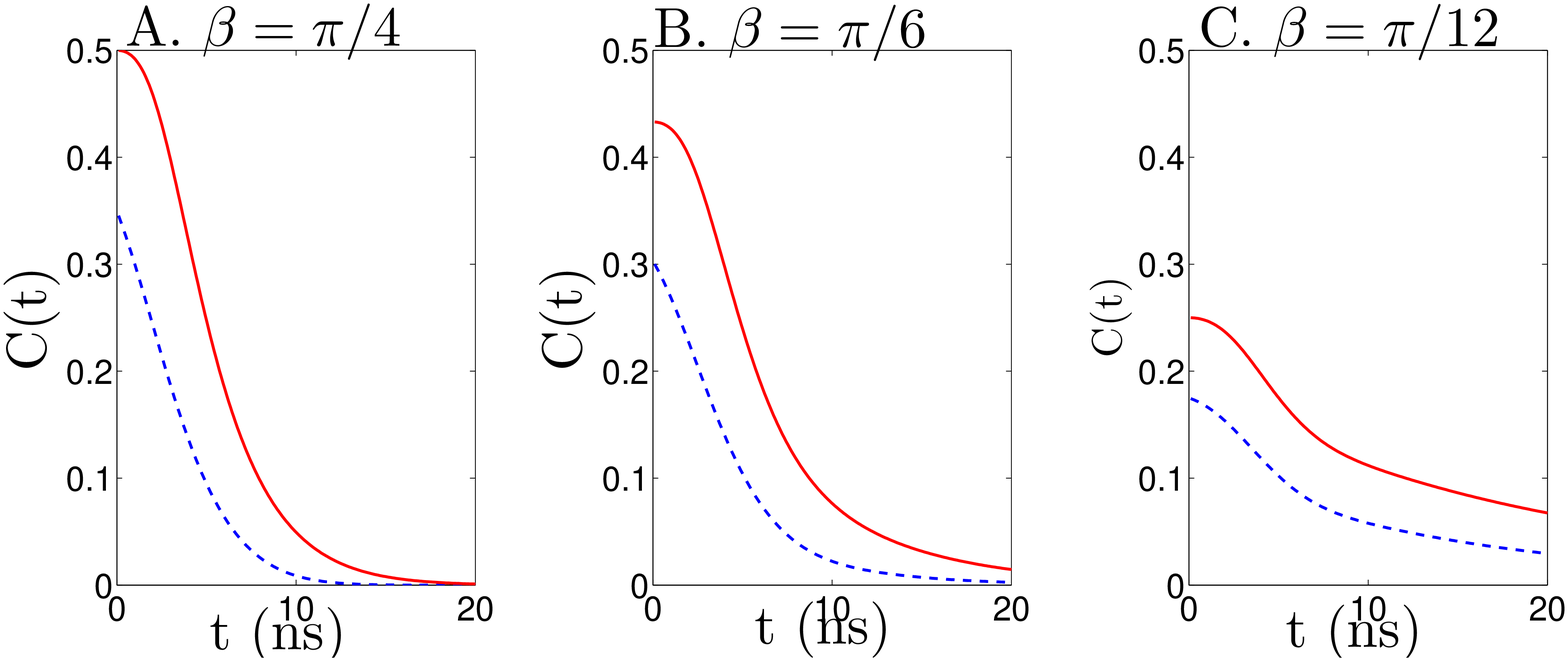}
	\renewcommand{\figurename}{Fig.}
	\caption{Evolution of the coherence ratio $C(t)=\frac{|\rho_{e_1,e_2}|}{\rho_{e_1,e_1}+\rho_{e_2,e_2}}$ of a critically damped V-system ($\zeta \ll 1$). Here $\gbar=1.0$ and $\Delta=1.0$, but with varying pumping rate asymmetry $\gamma_1=\gbar \sin^2(\beta)$ and $\gamma_2=\gbar \cos^2(\beta)$ and dipole alignment, $p=1.0$ (solid) and $p=0.7$ (dashed).}
\label{fig:6}
\end{figure}

In this case, the coherence ratio, $C(t) =\frac{|\rho_{e_1e_2}|}{|\rho_{e_1e_1}\rho_{e_2e_2}|}$, is, when $\gamma_1=\gamma_2$
\beq
C(t)=\frac{p\gamma}{2}\left(\frac{e^{-\gamma t}t\sqrt{1+\Delta^2t^2/4}}{1-e^{-\gamma t}+\Delta^2t^2/2e^{-\gamma t}}\right).
\label{eq:CDC}
\eeq
The ratio $C(t)$ in the critical region is shown in \cref{fig:6}.
\Cref{eq:CDC} shows the same $p$ scaling as in the other regimes.
However, the dependence on $\gamma_i$ is hidden due to the $\gamma_1=\gamma_2$ assumption.
\Cref{fig:6} shows that the same suppression of $C(t)$ with increasing asymmetry as in the underdamped (\cref{fig:2}) and overdamped(\cref{fig:4}) regions.

\section{Conclusion}
\label{sec:Conc}
We have presented analytical solutions for the dynamics of a V-system interacting with an incoherent radiation bath in three limiting cases.
In all three of these regimes the dynamics can be understood through the interplay of interference effects and radiative decay processes.
Our results show a transient suppression of population in the presence of coherences relative to rate law dynamics in the overdamped and critically damped regimes.
These findings indicate that the in-phase superposition, $\ket{\phi_+}$ prepared in the excited state manifold by incoherent light excitation shows constructive interference of radiative decay processes, leading to the reduced population of the excited state manifold.
Coherent effects are transient in all regimes (with the exception of the $\Delta_p=0$ case) and the system eventually collapses into the incoherent mixtures predicted by Pauli rate-law considerations.
However, in the overdamped region ($\zeta={\gbar}/{\Delta_p} \gg 1$) the coherent superposition $\ket{\phi_+}$ is remarkably long-lived and survives for $\tau_{long}={2\gbar}/{(\Delta_p^2)}$.

A striking feature of the Bloch-Redfield Master Equations [\cref{eq:QME}] is that incoherent light excitation produces mixtures, $\rho_d$ [\cref{eq:rhod}], in the excited state manifold that contain in-phase coherent superpositions of the excited states, $\ket{\phi_+}=\frac{1}{\sqrt{2\rbar}}(r_1\ket{e_1}+r_2\ket{e_2})$.
In-phase superpositions, such as $\ket{\phi_+}$, lead to constructive interference in the emission process, thereby suppressing the population of the excited state manifold as seen in the overdamped and critically damped regimes [\cref{eq:ODP,eq:CDP,fig:3,fig:5}].
This observation, however, does not contradict the previously reported enhancement of photocell currents due to coherences between the excited states \cite{Scully11,Scully13} as the photocell systems contain an additional conduction-band reservoir state $|\alpha\rangle$ to which the two excited states decay, and to which
the coherent enhancement of decay also applies.
The observed enhancement of photocurrent in these systems therefore suggests that, rather than enhancing absorption or suppressing the radiative decay to the ground state, the coherent enhancement of the decay rate to the conduction-band reservoir state $|\alpha\rangle$ outweighs the increased decay rate to the ground state.

Our findings motivate further consideration of the incoherent-light excitation of multilevel systems.
First, the variation of parameters method used here allows for the consideration of time-dependent radiation fields.
The investigation of adiabatic turn-on of the radiation field \cite{Grinev,amr2} is particularly important in the context of natural (sunlight) excitation of biomolecules and LHC's, since 
the natural  turn-on timescales are far longer than typically exploited in femtosecond experiments.
Second, the Bloch-Redfield method employed here predicts rich dynamics in systems with more complicated ground state manifolds, where the non-secular terms produce additional phenomena due to interference effects between different ground states. Work on these extensions is in progress.

\begin{acknowledgements}
This work was supported by the US AFOSR through contact number FA9550-13-1-0005, and by NSERC.
\end{acknowledgements}
\appendix
\section{Derivation of V System Dynamics}

\subsection{Coefficient Matrix Spectrum}
\label{app:spec}
Consider first the coefficient matrix, A, in the ($\nbar \ll 1$) weak pumping limit.
\Cref{eq:Amat} can be rewritten as a simpler matrix perturbed by a term of order $\nbar \ll1$.
\beq
A=
\left(
\begin{array}{cccc}
-\gamma_1 & 0 & -p\sqrt{\gamma_1\gamma_2} & 0\\
0 & -\gamma_2 & -p\sqrt{\gamma_1\gamma_2} & 0\\
-\frac{p}{2}\sqrt{\gamma_1\gamma_2} & -\frac{p}{2}\sqrt{\gamma_1\gamma_2} & -\gbar & \Delta\\
0 & 0 & -\Delta & -\gbar 
\end{array}
\right)
+
\nbar
\left(
\begin{array}{cccc}
-\gamma_1 & -\gamma_1 & -p\sqrt{\gamma_1\gamma_2} & 0\\
-\gamma_2 & -\gamma_2 & -p\sqrt{\gamma_1\gamma_2} & 0\\
-\frac{3p}{2}\sqrt{\gamma_1\gamma_2} & -\frac{3p}{2}\sqrt{\gamma_1\gamma_2} & -\gbar & 0\\
0 & 0 & 0 & -\gbar
\end{array}
\right)
=A^{(0)} +\nbar A'
\label{eq:Apert}
\eeq
where $\gbar =\half(\gamma_1+\gamma_2)$ and $\rbar=\half(r_1+r_2)$ are the arithmetic mean decay widths and pumping rate respectively.
We treat the perturbative expansion of $A$ given by \cref{eq:Apert} to zeroth order in $\nbar \ll 1$.
The results obtained to zeroth order in $\nbar$ are in extremely close agreement with the numerically exact solutions and clearly illustrate the physics of the system.

Consider now the spectrum, $\{\lambda_i\}$, of $A^{(0)}$ to obtain \cref{eq:lambda} in the main text.\
After some elementary manipulations of $\det(A^{(0)}-\lambda I)$, the characteristic polynomial of $A^{(0)}$ takes the form
\beq
\det(A^{(0)}-\lambda I)= (\gamma_1+\lambda)(\gamma_2+\lambda)[(\gbar+\lambda)^2+\Delta^2]-p^2\gamma_1\gamma_2(\gbar+\lambda)^2
\label{eq:CharPoly}
\eeq

\Cref{eq:CharPoly} can be rewritten in the following biquadratic form:
\beq
\det(A^{(0)}-\lambda I)= x^2 + [\gbar^2 +(1-p^2)\gamma_1\gamma_2 +\Delta^2]x +[\gbar^2(1-p^2)\gamma_1\gamma_2 +\gamma_1\gamma_2\Delta^2]
\label{eq:biquad}
\eeq
where $x=\lambda(\lambda+2\gbar)$.
Applying the quadratic formula, first to \cref{eq:biquad}, to obtain $x$ then to $\lambda^2+2\gbar\lambda-x=0$ gives the spectrum of $A^{(0)}$.
\beqs
\beq
\lambda_i =-\gbar \pm \sqrt{\gbar^2+x_\pm}
\label{eq:lamrough}%
\eeq
\beq
x_\pm =-\half a_1 \pm \half\sqrt{a_1^2-4a_0}
\label{eq:x}%
\eeq
\label{eq:Aspec}%
\eeqs
where $a_1 = \gbar^2+(1-p^2)\gamma_1\gamma_2 +\Delta^2$ and $a_0=\gamma_1\gamma_2(\Delta^2+(1-p^2)\gbar^2)$ are the coefficients of the $x^1$ and $x^0$ terms of \cref{eq:biquad} respectively.
Simplifying \cref{eq:Aspec} and substituting $\Delta_p=\sqrt{\Delta^2+(1-p^2)\gamma_1\gamma_2}$ yields \cref{eq:lambda} in the main text:

\beq
\lambda_i=-\gbar \pm \sqrt{\frac{(\gbar^2-\Delta_p^2)\pm \sqrt{(\gbar^2-\Delta_p^2)^2+\Delta^2(\gamma_1-\gamma_2)^2}}{2}}
\label{eq:lambda2}
\eeq

\Cref{eq:lambda2} can be rewritten in the following convenient form
\beq
\lambda_i=-\gbar \pm \Delta_p\sqrt{\zeta^2-1}\sqrt{\frac{1\pm\sqrt{1+\eta^2}}{2}}
\label{eq:lamDO2}
\eeq
where $\zeta =\frac{\gbar}{\Delta_p}$ and $\eta=\frac{\Delta|\gamma_1-\gamma_2|}{|\gbar^2-\Delta_p^2|}$ are parameters defined in \cref{eq:zeta,eq:eta} of the main text respectively.

\subsection{Underdamped Regime $\zeta =\frac{\gbar}{\Delta_p} \ll 1$}
\label{app:UDamp}
Consider now a V-system in the underdamped regime with $\zeta \ll1$.
In such a system, the constraint on $\zeta$ is equivalent to
\beq
\Delta_p^2 =\Delta^2 +(1-p^2)\gamma_1\gamma_2 \gg \gbar^2
\label{eq:UDCond}
\eeq

In \cref{eq:UDCond}, ${\gamma_1\gamma_2}$ is the square of the geometric mean while $\gbar^2$ is the square of the arithmetic mean.
The geometric mean is less than or equal to the arithmetic mean and so $\gamma_1\gamma_2\leq\gbar^2$.
Noting that $0\leq p\leq1$, \cref{eq:UDCond} gives the conditions for the underdamped regime

\beqs
\beq
 \Delta \gg \gamma_i
\eeq
\beq 
\Delta_p \approx\Delta
\eeq
\label{eq:UDCondg}
\eeqs

We now consider $\eta =\frac{\Delta|\gamma_1-\gamma_2|}{|\Delta_p^2-\gbar^2|}$.
Applying the conditions in \cref{eq:UDCondg}, we get

\beq
\eta \approx \frac{|\gamma_1-\gamma_2|}{\Delta} \ll 1
\label{eq:UDetacond}
\eeq

We can now apply \cref{eq:UDetacond} and the $\zeta \ll 1$ condition to approximate \cref{eq:lamDO2} term by term using the binomial approximation. This gives:

\beqs
\beq
\sqrt{\zeta^2-1}\approx i(1-\frac{\zeta^2}{2}) \approx i
\eeq
\beq
\sqrt{\frac{1 \pm\sqrt{1+\eta^2}}{2}} \approx 
\begin{cases}
\sqrt{1+\frac{\eta^2}{4}} \approx 1 &  \\
i\frac{\eta}{2} &
\end{cases}
\label{eq:etaBi}
\eeq
\label{eq:UDBi}
\eeqs

Substituting \cref{eq:UDBi} into \cref{eq:lamDO2} gives the spectrum, $\{\lambda_i\}$, of $A$ in the underdamped regime

\beqs
\beq
\lambda_1 = -\gamma_1
\label{eq:lam1}
\eeq
\beq
\lambda_2 = -\gamma_2
\label{eq:lam2}
\eeq
\beq
\lambda_{3,4} = -\gbar \pm i\Delta_p
\label{eq:lam34}
\eeq
\label{eq:lams}
\eeqs

Now we proceed to determine the normal modes of $A^{(0)}$ (\cref{eq:Apert}).
The eigenvectors, $\mathbf{v}_i \in null(A^{(0)}-\lambda_i I)$, can be found with the usual method to give

\beqs
\beq
\mathbf{v}_1 \propto [1,0,0,\frac{p\sqrt{\gamma_1\gamma_2}}{2\Delta_p}]
\label{eq:v1}
\eeq
\beq
\mathbf{v}_2 \propto [0,1,0,\frac{p\sqrt{\gamma_1\gamma_2}}{2\Delta_p}]
\label{eq:v2}
\eeq
\beq
\mathbf{v}_3 \propto [0,0,1,1]
\label{eq:v3}
\eeq
\beq
\mathbf{v}_4 \propto [0,0,1,-1]
\label{eq:v4}
\eeq
\label{eq:vs}
\eeqs

We now use the general variation of parameters method \cite{BDP} with the initial conditions, $\mathbf{x_0}=\mathbf{0}$, appropriate for excitation from the ground state to get the general form of the weak solutions of the master equations \cref{eq:QME}

\beq
\mathbf{x}(t)=e^{At}\mathbf{x_0} +  \int^t_0 ds e^{A(t-s)}\mathbf{d} \to \sum_{i=1}^4 \int^t_0 ds (\mathbf{v}_i\cdot\mathbf{d})e^{\lambda_i (t-s)} \mathbf{v}_i
\label{eq:GenSol2}
\eeq

Substituting \cref{eq:lams,eq:vs} into \cref{eq:GenSol2}, evaluating the integrals gives, to lowest contributing order in $\zeta$ the density matrix elements

\beqs
\beq
\rho_{e_i,e_i}(t)=\nbar (1-e^{-\gamma_it})
\label{eq:UDP2}
\eeq
\beq
\rho_{e_1,e_2}^R(t)=\frac{p \sqrt{r_1r_2}}{\Delta_p}e^{-\gbar t}\sin(\Delta_p t)
\label{eq:UDCR2}
\eeq
\beq
\rho_{e_1,e_2}^I(t)=\frac{p \sqrt{r_1r_2}}{\Delta_p}\left( e^{-\gbar t}(\cos(\Delta_p t)-1)-\frac{e^{-\gamma_1 t}-e^{-\gamma_2t}}{2} \right)
\label{eq:UDCI2}
\eeq
\label{eq:UDamp2}
\eeqs

\Cref{eq:UDamp2} gives the result presented in the main text. 

\subsection{Overdamped Regime $\zeta \gg 1$}
\label{app:ODamp}
Consider now the overdamped case, characterized by $\zeta \gg 1$.
This is equivalent to $\gbar \gg \Delta_p$.
As in the underdamped case we consider the parameter $\eta$

\beq
\eta = \frac{\Delta|\gamma_1-\gamma_2|}{|\gbar^2-\Delta_p^2|} \approx \frac{\Delta|\gamma_1-\gamma_2|}{\gbar^2} \leq \frac{2\Delta_p}{\gbar} \ll 1
\label{eq:ODampEta}
\eeq

We now apply a term by term binomial approximation to \cref{eq:lamDO2}, this time using $\zeta^{-1} \ll 1$

\beq
\sqrt{1-\zeta^{-2}} \approx 1-\frac{1}{2\zeta^2}
\label{eq:ODBi}
\eeq

Substituting the binomial approximations in $\zeta^{-1}$ (\cref{eq:ODBi}) and in $\eta$ (\cref{eq:etaBi}) into \cref{eq:lamDO2} we obtain the spectrum of $A^{(0)}$ in the overdamped region.

\beqs
\beq
\label{eq:Olam1}
\lambda_1=-2\gbar
\eeq
\beq
\lambda_2=-\frac{\Delta_p^2}{2\gbar}
\label{eq:Olam2}
\eeq
\beq
\lambda_{2,3}=-\gbar(1 \pm i\frac{\eta}{2}) \approx -\gbar
\label{eq:Olam34}
\eeq
\label{eq:Olams}
\eeqs

Proceeding to find the eigenvectors of $A^{(0)}$ corresponding to the spectrum [\cref{eq:Olams}] through the standard method:
\beqs
\beq
\mathbf{v}_1 \propto [r_1,r_2,p\sqrt{r_1r_2},0]
\label{eq:Ov1}%
\eeq
\beq
\mathbf{v}_2 \propto [r_2,r_1,-p\sqrt{r_1r_2},0]
\label{eq:Ov2}%
\eeq
\beq
\mathbf{v}_3 \propto [0,0,0,1]
\label{eq:Ov3}%
\eeq
\beq
\mathbf{v}_4 \propto [1,-1,-\frac{\gamma_1-\gamma_2}{p\sqrt{\gamma_1\gamma_2}},0]
\label{eq:Ov4}%
\eeq
\label{eq:Ovs}%
\eeqs
We note here that $\mathbf{v}_1 =\mathbf{d}$ (\cref{eq:Ov1}) indicating that incoherent pumping drives V-type systems into the statistical mixture $\rho_d$ discussed in the main text (\cref{eq:rhod}).
In contrast, $\mathbf{v_2}$ (\cref{eq:Ov2}) represents the decay from $\rho_d$ to the rate-law predicted mixture $\rho_{eq}$ (\cref{eq:rhoeq}).

Finally, substituting \cref{eq:Olams,eq:Ovs} into \cref{eq:GenSol2} and computing the integrals we obtain the density matrix elements

\beqs
\beq
\rho_{e_i,e_i}(t)=\frac{1}{2\gbar}\left[r_i(1-e^{-2\gbar t})+r_{j\neq i}(1-e^{-\frac{\Delta_p^2}{2\gbar}t}) \right]
\label{eq:ODP2}
\eeq
\beq
\rho_{e_1,e_2}^R(t)=\frac{p \sqrt{r_1r_2}}{2\gbar}\left[e^{-\frac{\Delta_p^2}{2\gbar}t}-e^{-2\gbar t}\right]
\label{eq:ODCR2}
\eeq
\label{eq:ODamp2}
\eeqs

as discussed in the main text.

\subsection{Critically Damped Regime $ \zeta = 1$; $\eta = \frac{\Delta|\gamma_1-\gamma_2|}{|\Delta_p^2 -\gbar^2|}=0$}
\label{app:CDamp}
Consider the final regime discussed in the main text, the critical regime ($\zeta=1$ and $\eta=0$).
In this case, the spectrum, $\{\lambda_i \}$, (\cref{eq:lamDO2}) collapses to a single eigenvalue $\lambda =-\gbar$.

Note that $\eta = 0$ can occur in two cases, if $\Delta = 0$ or if $\gamma_1=\gamma_2$.
Consider first, the case of $\Delta =0$.
In this case $\zeta =1$ (and so $\Delta_p=\gbar$) implies that $p=1$ and $\gamma_1=\gamma_2$.
Substituting this back into the equation $\Delta_p=\gbar$ shows that this is simply the trivial case of uncoupled states ($\gamma_1=0=\gamma_2$).
Consider next, the $\gamma_1=\gamma=\gamma_2$ case.
In this case, $\zeta=1$ implies that $\Delta=p\gamma$.

In the critically damped regime, the matrix $A^{(0)}$ becomes defective and cannot be diagonalized.
The variation of parameters procedure still provides the solution to the Quantum master Equations \cref{eq:QME} in the form.

\beq
\mathbf{x}(t)= \int^t_0 ds e^{A(t-s)}\mathbf{d}
\label{eq:GenSolJCF}
\eeq

Where the exponential of $A$ can be found by putting $A$ into Jordan Canonical Form.
This can be done by determining the generalized eigenvectors.
Proceeding to find the generalized eigenvectors of rank $j$, $\mathbf{v}_i^{(j)}$, using the standard procedure we obtain

\beqs
\beq
\mathbf{v}_1^{(1)} \propto [1, -1,0,0]
\label{eq:Cv1}
\eeq
\beq
\mathbf{v}_2^{(1)} \propto [1, 1,0,-1]
\label{eq:Cv21}
\eeq
\beq
\mathbf{v}_2^{(2)} \propto [0, 0, \Delta^{-1},0]
\label{eq:Cv22}
\eeq
\beq
\mathbf{v}_2^{(3)} \propto [0, 0,0,\Delta^{-2}]
\label{eq:Cv33}
\eeq
\label{eq:Cvs}
\eeqs

In the basis given by \cref{eq:Cvs} the coefficient matrix takes the Jordan Canonical Form

\beq
A=\left(
\begin{array}{cccc}
-\gamma & 0 & 0& 0 \\
0 & -\gamma & 1& 0 \\
0& 0& -\gamma & 1\\
0& 0& 0 & -\gamma
\end{array}
\right)
\label{eq:JCFA}
\eeq

Finally, substituting \cref{eq:Cvs,eq:JCFA} into \cref{eq:GenSolJCF} and evaluating the integrals we obtain the dynamics of a V-system in the critical regime

\beqs
\beq
\rho_{e_i,e_i}(t)=\nbar (1-e^{-\gamma t}+\frac{p^2\gamma^2}{2}t^2e^{-\gamma t})
\label{eq:CDP2}
\eeq
\beq
\rho_{e_1,e_2}^R(t)=\nbar p\gamma t e^{-\gamma t}
\label{eq:CDCR2}
\eeq
\beq
\rho_{e_1,e_2}^I(t)=-\nbar \frac{p^2\gamma^2}{2}t^2 e^{-\gamma t}
\label{eq:CDCI2}
\eeq
\label{eq:CDamp2}
\eeqs

this reproduces the dynamics discussed in the main text.


\begin{thebibliography}{99}
\bibitem{note1}
Note that while Fig. 3 of Ref. \cite{Agarwal01} shows the absolute magnitude of the coherence vs. time, the time evolution is not discussed in that paper, which focuses on the coherences in the long-time limit.
\end{thebibliography}
\end{document}